\documentclass[11pt]{article}
\begin{document}
\def\refs#1{\ref{#1}.}    
\def\refe#1{(\ref{#1})}
\def\beq{\begin{equation}}
\def\beql#1{\begin{equation}\label{#1}}
\def\eeq{\end{equation}}
\def\bay{\begin{eqnarray}}
\def\bayl#1{\begin{eqnarray}\label{#1}}
\def\eay{\end{eqnarray}}
\def\b{\beta}
\def\r{\rho}
\def\s{\sigma}
\def\l{\lambda}
\def\D{\Delta}
\def\ln{\mbox{ln}}
\def\e{\eta}
\def\t{\tau}
\def\pb{{\bar p}}

\title{Shannon information increase and rescue in friction}
\author{Lajos Di\'osi\\
Research Institute for Particle and Nuclear Physics\\
H-1525 Budapest 114, POB 49, Hungary}
\maketitle

\begin{abstract}
On the standard microscopic model of friction we confirm the common belief
that the irreversible entropy production originates from the increase of 
Shannon information. We reveal that the reversible microscopic dynamics
would continuously violate the Gibbsian interchangeability of molecules. 
The spontaneous restoration of interchangeability constitutes the mechanism 
of irreversibility. This is the main suggestion of the Letter. We also point 
out that the irreversible entropy can partially be rescued if the size of the 
medium is finite. The rescueable entropy is found to be proportional to the 
square of the Fisher statistical (also called thermodynamic) distance.
\end{abstract}

{\it Introduction.\/} It is commonly trusted that the thermodynamic entropy 
produced in irreversible processes is identical to the increase of the 
Shannon information in the underlying microdynamic process. A direct proof 
is yet missing. The microscopic dynamics is reversible, it preserves the 
Shannon information. We need a transparent model to resolve this fundamental 
paradox. In this Letter I discuss the mechanical friction of a thin `disc' 
perpendicularly driven through a medium with small velocity $V$. The 
phenomenological model is a prototype for the whole theory of irreversible 
processes \cite{LanLif}. Accordingly, the disc moves against the frictional 
force $-\e V$ where $\e$ is the friction constant. If the medium (reservoir) 
is in thermal equilibrium at inverse temperature $\b$ then the rate of 
thermodynamic entropy production can be written in this form:
\beql{dotS}
\dot S=\e\b V^2~.
\eeq
An elementary microscopic model will be considered where the medium is ideal 
gas of light molecules scattering elastically on both flat surfaces of the 
disc. This is the Brownian motion microscopic model well-known from 
textbooks \cite{Isi}, applied this time to a disc instead of a `ball' 
\cite{ball}. The model enables us to calculate the exact {\it reversible} 
evolution of the microscopic system. I will confirm that the thermodynamic 
entropy production rate \refe{dotS} can indeed be identified as the rate 
$\dot I$ of Shannon information increase in an infinite medium:
\beql{IS}
\dot I = \dot S~,
\eeq
if the molecular interchangeability, which is being violated by each 
reversible collision on the disc, is being re-imposed ({\it \`a la} Gibbs). 
This appears to be a generic mechanism of irreversibility in the otherwise 
reversible dynamics. 

The molecular interchangeability is less relevant if the medium is 
{\it finite}. Then, as we shall prove, the increase of Shannon information 
becomes less than the standard irreversible entropy production \refe{dotS}:
\beql{dotI}
\dot I = \dot S - \dot S_{resc}~.
\eeq
In the given microscopic model, the rate of the {\it rescued} information 
(and entropy) will take the following form:
\beql{Sresc}
\dot S_{resc}\equiv\dot I_{resc}=\chi\dot  S~,
\eeq
where $\dot S$ is the standard irreversible entropy production rate 
\refe{dotS} and the rescue factor $\chi$ is the ratio of the already 
interacted molecules within the whole reservoir.

{\it Shannon information.\/} Let the medium be ideal gas of $N$ identical 
molecules of mass $m$. The initial distribution of the momenta 
$p_1,\dots,p_N$ is the thermal equilibrium distribution at inverse 
temperature $\b$: 
\beql{r0}
\r_0(p_1,\dots,p_N)=\exp\left[-\frac{\b}{2m}\sum_{r=1}^N p_r^2\right]~.
\eeq
The molecules collide on the disc elastically (reversibly). The mean 
intercollision time $\t$ is assumed to be independent of $V$ which is a good 
approximation for small velocities $V$. In a single collision, the momentum 
change of the given molecule is approximately the following:
\beql{Dp1}
p_r \rightarrow  2mV - 2p_r~,
\eeq
provided the molecules are much lighter than the disc. In a short yet 
macroscopic interval $\D t\gg\t$, the average momentum transfer to the medium 
is $2mV\D t/\t$. Therefore the average frictional force is $-2mV/\t$ while  
the friction constant takes this simple form:
\beql{eta}
\eta=\frac{2m}{\t}~.
\eeq
(For simplicity, we request that no molecules will scatter twice or more. 
It also means that the mass of the medium is being {\it consumed} by the 
collisions at rate $\e/2$. Then, in a finite reservoir of mass $M_R=Nm$,
the friction process would be cut at time $2M_R/\e$.)

After time $\D t$, the equilibrium distribution is shifted:
\beql{rDt}
\r_{\D t}(p_1,\dots,p_N)=\r_0(p_1-\pb_1,\dots,p_N-\pb_N)~,
\eeq
where $\pb_r$ is the mean momentum transfer caused by collision to the 
r'th molecule. Obviously, the Shannon information has been preserved by 
the collisions. On the other hand, we know from the thermodynamics of
friction that {\it there is} a definite entropy production \refe{dotS}. 
We suspect that an ultimate information increase comes from the
undistinguishability of the molecules. We assume that knowledge
is leaking out, regarding {\it which molecule scattered on the disc and 
which did not}. Accordingly, the true distribution $\r_{\D t}$ of the 
medium must be the re-symmetrized projection of the post-collision 
distribution \refe{rDt} over all permutations $\pi$ of the $N$ molecules:  
\beql{rDtsim}
\r_{\D t}(p_1,\dots,p_N)=
\frac{1}{N!}\sum_{\pi}\r_0(p_{\pi(1)}-\pb_1,\dots,p_{\pi(N)}-\pb_N)~.
\eeq
We define the increment of the Shannon information (counted in $\ln2$ bit 
units) during the interval $\D t$:
\beql{DI}
\D I = -\int\left[\r_{\D t}\ln\r_{\D t}-\r_0\ln\r_0\right]
\frac{d^Np}{(2\pi m\b)^{N/2}}~.
\eeq
Using the Eqs.(\ref{r0},\ref{rDtsim}), I have derived the Taylor-expansion
upto the lowest non-vanishing order of the momentum transfers:
\beql{DI1}
\D I=\frac{\b}{2m}
\left[\sum_{r=1}^N\pb_r^2-\frac{1}{N}(\sum_{r=1}^N\pb_r)^2\right]~.
\eeq
The two terms on the R.H.S. correspond to macroscopic quantities.
There are two relationships, respectively. The first is the balance
between the average energy increase of the medium and the energy
dissipated by the friction force:
\beql{ene} 
\frac{1}{2m}
\sum_{r=1}^N\pb_r^2= \e V \times V\D t~.
\eeq
Similarly, there is an average momentum balance as well:
\beql{mom} 
\sum_{r=1}^N\pb_r= \e V \times \D t~.
\eeq
Let us insert the above two relationships into \refe{DI1}, yielding
the following rate of information increase:
\beql{dotI1}
\frac{\D I}{\D t}=\e\b V^2\left(1-\frac{\e\D t}{2M_R}\right)~.
\eeq
This is the central result of the Letter. It confirms the equality 
\refe{IS} of irreversible information \refe{dotI1} and entropy 
\refe{dotS} productions in {\it infinite} mass reservoirs $(M_R=Nm=\infty)$.

{\it Information rescue.\/} In a finite reservoir, the Eq.\refe{dotI1} 
contains a non-stationary correction, we call it the {\it rescued} 
information. Its rate reads:
\beql{Iresc}
\frac{\D I_{resc}}{\D t}=\e\b V^2\frac{\e\D t}{2M_R}~.
\eeq
This equation is equivalent with the Eq.\refe{Sresc} if we set 
$\chi=\frac{\e\D t}{2M_R}$ 
which is indeed the ratio of the consumed mass of molecules to the
total mass of the reservoir. In conventional system--reservoir setups the 
mass $M_R$ of the reservoir is kept constant. Yet we can consider 
non-static setups where the reservoir mass is allocated dynamically, i.e., 
varying with time. We shall therefore introduce a monotone function 
$M_R(t)$ with fixed initial $M_R(0)=0$ and final values 
$M_R(t_{tot})=M_{Rtot}$, where $t_{tot}$ is the total considered time of the 
process.
(The allocation rate $\dot M_R$ should never fall below the consumption rate
$\dot M_R\geq\e/2$. In the rest of the Letter we shall understood
this condition, without mentioning it again.)
The function $M_R(t)$ will govern the gradual {\it allocation} of the medium 
to the friction process. A discretized interpretation is the following. 
We divide the total time and reservoir into small yet macroscopic parts. 
In each short interval 
\mbox{$[t,t+\D t]$} in turn, a small part
\mbox{$\D M_R(t)=M_R(t+\D t)-M_R(t)$}
of the total mass $M_{Rtot}$ is used as medium: the small medium is 
brought into contact with the disc at the beginning of the small period and 
removed at the end of it, to be replaced by the forthcoming small medium
{\it e.t.c.}, until the whole period $[0,t_{tot}]$ and the total 
reservoir mass $M_{Rtot}$ are both exhausted. To the first short period 
$[0,\D t]$, in particular, we allocate a small mass 
\mbox{$M_R(\D t)\approx\dot M_R(0)\D t$}. Substituting 
$M_R$ by $\dot M_R(0)\D t$, the rate \refe{Iresc} of the 
rescued information becomes stationary on the macroscopic time scale:
\beql{dotIresc}
\dot I_{resc}=\e\b V^2 \frac{\e}{2\dot M_R}~.
\eeq
We can extend this differential relationship for all later time. 
Invoking the Eq.\refe{Sresc}, we can read out the rescue factor
for the case of time-dependent reservoir allocation: 
\beql{chi}
\chi=\frac{\e}{2\dot M_R}~.
\eeq
This is the ratio of the reservoir's consumption rate $\e/2$ with respect to 
its allocation rate $\dot M_R$. 

An optimization of the rescued information (entropy) is 
straightforward. We are going to maximize the total rescued information:
\beql{Iresctot}
I_{resc}\equiv S_{resc}=\e\b
\int_0^{t_{tot}} V^2 \frac{\e}{2\dot M_R}dt~,
\eeq
by varying $M_R(t)$ at fixed total reservoir mass $M_{Rtot}$. The 
Euler-Lagrange equation yields $\dot M_R=\mbox{const}\times V$. 
The optimum reservoir allocation $M_R(t)$ is thus proportional to the
length $\int_0^t Vdt$ of the disc trajectory. We should allocate equal 
reservoir masses over equal lengths of the driven trajectory of the disc. 
For the special case of stationary driving $V=\mbox{const}$, the optimum 
allocation is also stationary: $\dot M_R\equiv M_{Rtot}/t_{tot}$.
The optimum ratio \refe{chi} of the rescued/standard information (entropy) 
takes this form:
\beql{chiopt}
\chi=\frac{\e t_{tot}}{2M_{Rtot}}~.
\eeq

{\it Reservoir-driven process.} We apply our results to the case when
we do not directly drive the disc. Rather we change slowly the velocity
$V_R(t)$ of the medium and only the frictional force $-\e(V-V_R)$ will 
drive the disc. All previous equations remain valid if 
$V$ is substituted by $V-V_R$ which can be further expressed from the
Newton equation $M\dot V=-\e(V-V_R)$ where $M$ is the disc mass.
The resulting substitution has thus to be
$V\rightarrow -\frac{M}{\e}\dot V$.
The thermodynamic entropy production rate \refe{dotS} takes this form:
\beql{dotS1}
\dot S=\b\frac{M^2}{\e}\dot V^2~,
\eeq
and for the rescued information (entropy) rate \refe{dotIresc} we obtain:
\beql{dotIresc1}
\dot I_{resc}=\b M^2\dot V^2 \frac{1}{2\dot M_R}~.
\eeq
It is remarkable that the friction constant $\e$ canceled! 
The rescued information in the reservoir-driven process is universally 
independent of the temporal scales. (The rescue factor $\chi$ will, 
however, keep its $\e$-dependence \refe{chi}.) The optimization of the 
rescued information (entropy) is again straightforward. We minimize the 
total rescued information:
\beql{Iresctot1}
I_{resc}\equiv S_{resc}=\b M^2
\int_0^{t_{tot}}\dot V^2 \frac{1}{2\dot M_R}dt
\eeq
at fixed total reservoir mass $M_{Rtot}$. The variation of the allocation
function $M_R(t)$ yields the condition 
$\dot M_R=\mbox{const}\times \dot V$. 
The optimum reservoir allocation $M_R(t)$ is proportional to the
velocity change $V(t)-V(0)$ of the disc. For the special case of constant 
acceleration $\dot V=\mbox{const}$, the optimum allocation is stationary: 
$\dot M_R\equiv M_{Rtot}/t_{tot}$. The optimum rescue factor is the same 
\refe{chiopt} as for the externally driven disc.

Finally we give an explicit interpretation for the `timelessness' of the
rescued information \refe{dotIresc1}. We note that the velocity $V$ of the 
reservoir--driven disc is an equilibrium thermodynamic variable itself. 
It is tending to equilibrate with the medium's velocity $V_R$. 
In equilibrium it has a definite fluctuation $\s_V$ which is also a 
thermodynamic quantity: $\s_V^2=1/\b M$. This gives rise to the notion of 
{\it statistical length} $\ell$ \cite{FisWooBraCav}, also called 
thermodynamic length \cite{WeiRupDio}:  
\beql{ell}
d\ell=\frac{dV}{\s_V}~,
\eeq
rescaling the parameter length $dV$ in absolute units of equilibrium
thermodynamics. Let us express the evolution in function of the statistical 
length $\ell$ instead of the time $t$. Replacing \refe{dotIresc1}, 
we obtain the following result for the rescued entropy rate:
\beql{primeIresc1}
I_{resc}'\equiv S_{resc}'=\frac{M}{2M_R'}~,
\eeq
where the apostrophes stand for $\ell$-derivatives.
The maximum of the rescued entropy is obtained by variation of the 
reservoir allocation function $M_R(\ell)$ between the fixed endpoints.
The resulting optimum is $M_R(\ell)=\mbox{const}\times \ell$.
In the thermodynamic length, the optimum rescued entropy
rate \refe{primeIresc1} is constant and the total rescued entropy
becomes a simple quadratic function of the total statistical length:
\beql{Iresctot1ell}
I_{resc}\equiv S_{resc}=\frac{M}{2M_{Rtot}}\ell_{tot}^2~.
\eeq
It depends exclusively on the {\it equilibrium} thermodynamics of the 
disc--reservoir frictional interaction. 

{\it Discussion.} On the common microscopic model of friction we have 
pointed out that the standard thermodynamic entropy production derives 
exactly from the increase of Shannon information. The well-known 
discrepancy with the microscopic reversibility has been resolved in a 
natural way. The resolution comes from a related discrepancy, revealed 
by the Letter: The microscopic model is inevitably violating the Gibbsian 
interchangeability of molecules. The recovery, i.e., the re-symmetrization 
of the phase-space density is obviously irreversible. This is {\it the} 
mechanism of Shannon information increase and of macroscopic entropy 
production. (The role of interchangeability in the irreversibility of 
system-reservoir interactions has independently been noticed for a quantum 
informatic model of non-thermal equilibration by Ziman {\it et al.} 
\cite{Zimetal02}.) The first proof of perfect coincidence between the 
Shannon information and the phenomenological entropy production in a common 
irreversible process has been achieved by the present Letter. 
The phenomenology of friction is structurally identical to the model of 
general irreversible processes \cite{LanLif}. We therefore incline to think 
that our results reflects general features of irreversible processes.

We have also shown that the information increase falls with the size of
the medium. A certain portion of the standard entropy production can
thus be rescued in finite reservoirs. The Gibbsian molecular 
interchangeability, having been violated locally, will be recovered by a 
process propagating in space. The identity of the distinguished molecules 
(i.e. of those which `took part' in friction) will be smashed gradually 
over bigger and bigger parts of the surroundings. It is clear that this 
propagation is cut by the walls of the reservoir though temporal aspects 
(if any) of the propagation have not been discussed at all.

Entropy rescue might remain pure speculation for static reservoirs.
The Letter has discussed time-dependent reservoir allocations.
An externally driven `object' in a medium at rest as well as an object driven 
merely by the frictional force in a moving medium have been considered.
The optimum reservoir allocation is proportional to the real length of the 
object trajectory in the first case, and to the statistical (thermodynamical) 
length in the second case. The second result gives an exact microscopic 
interpretation for the role of thermodynamic length in irreversible process 
optimization. This may be a crucial step in the longstanding complex 
research  \cite{FisWooBraCav,WeiRupDio,SalBer83AndGor94SalNul98SalDio00}
to illuminate the significance of Fisher's statistical distance in 
statistics, control, thermodynamics, and quantum information.   

This work was supported by the Hungarian OTKA Grant 32640.


\begin{thebibliography}{99}

\bibitem{LanLif} L.D. Landau and E.M. Lifshitz, {\it Statistical Physics}
(Clarendon, Oxford, 1982).

\bibitem{Isi} 
A. Isihara, {\it Statistical Physics} (Academic Press, New York--London, 
         1971).

\bibitem{ball} Our present results can be generalized for the `ball'
at the additional cost (spared by the Letter) of using the joint phase-space 
distribution of the molecular momenta {\it and the coordinates} since the 
reversible description of the collisions needs the coordinates, too. 

\bibitem{FisWooBraCav}  
R.A. Fisher, Proc. R. Soc. Edinburgh {\bf 42}, 321 (1922).
W.K.Wootters, Phys. Rev. {\bf D23}, 357(1981);
S.L. Braunstein and C.M. Caves, Phys. Rev. Lett. {\bf 72}, 3439 (1994).

\bibitem{WeiRupDio} 
F.Weinhold, Phys. Today {\bf 29}, 23 (1976);
G.Ruppeiner, Phys.Rev. {\bf A20}, 1608 (1979);
L.Di\'osi, G.Forg\'acs, B.Luk\'acs, and H.L.Frisch, Phys.Rev. {\bf A29}, 
        3343 (1984).

\bibitem{Zimetal02} M.Ziman {\it et al.}, Phys. Rev. {\bf A65} 402105 (2002).

\bibitem{SalBer83AndGor94SalNul98SalDio00}
P. Salamon and R.S. Berry, Phys. Rev. Lett. 51 (1983) 1127;
B. Andresen and J.M. Gordon, Phys. Rev. {\bf E50}, 4346 (1994);
P.Salamon and J.D.Nulton, Europhys. Lett. {\bf 42} 571 (1998);
L.Di\'osi and P.Salamon, p286 in: {\it Thermodynamics of Energy Conversion
         and Transport}, eds.: S.Sieniutycz and A.deVos 
         (Springer, Berlin, 2000). 


\end{thebibliography}
\end{document}